# Power Saving Techniques for Wireless LANs

T. Simunic
CS Dept, UCSD
La Jolla, CA 92093
tajana@csl-mail.stanford.edu

**Abstract**

*Fast wireless access has rapidly become commonplace. Wireless access points and Hotspot servers are sprouting everywhere. Battery lifetime continues to be a critical issue in mobile computing. This paper first gives an overview of WLAN energy saving strategies, followed by an illustration of a system-level methodology for saving power in heterogeneous wireless environments.*

## 1. Overview of WLAN power saving techniques

Wireless communication today spans a wide range of devices, from cell phones to wireless sensors. Most of wireless data traffic is targeted at the infrastructure and is carried by wireless LAN (WLAN). Infrastructure and applications typically assume that devices are always connected resulting in a large battery drain on mobiles. Power management can make a big difference to the battery lifetime. A number of power reduction methodologies have been presented in the past covering all protocol layers, ranging from physical to the application layer. A short summary of different techniques is given in this paper while more details can be found in [[1],[2]].

Physical layer power optimization targets design of hardware components. Examples include minimizing the interconnect parasitic capacitance to reduce the dynamic power consumption and selectively turning off power supply to lessen leakage power. Power consumption of WLAN hardware is similar in transmit and receive modes [[2]]. Since WLANs spend as much as 90% of their time listening, the power control techniques aimed at reducing their transmission power are far from sufficient.

A number of power-saving MAC protocols have been proposed. 802.11 power saving standard has a device entering doze mode whenever there is no traffic for it in the traffic indication map sent by the access point. EC-MAC extends this by broadcasting a centrally determined schedule of data transmission times to reduce collisions and to provide exact times for entry into doze state. Longer mobile sleep periods can be created by aggregating MAC layer packets. Alternatively, with PAMAS nodes independently enter sleep state based on their battery levels.

Logical link layer formats packets and selects the error handling capabilities. Power savings are obtained by trading off retransmissions with Automatic Repeat Request (ARQ) with longer packet sizes due to Forward Error Correction. Adaptation of ARQ to the current channel state is another enhancement. Prediction of future channel conditions has a tradeoff on cost and the accuracy of prediction versus the energy savings given predicted conditions. Finally, a number of energy efficient ad-hoc routing protocols have been proposed.

Transport layer protocols are designed to work well when deployed on reliable links, thus causing problems when working in wireless conditions. This can be mitigated in various ways, ranging from splitting a connection, to probing, creating supporting links and completely new end-to-end protocols. Thus far standard TCP/IP and UDP implementations dominate wireless data communication.

At operating system level a number of techniques for controlling when wireless devices are on have been proposed in addition to more traditional CPU voltage scaling and scheduling. Decisions are made independently of any application information, and thus must relay on the quality of the predictive techniques.

Application level power optimization for WLANs can be grouped into three categories: load partitioning, proxy-based control and application-specific optimizations. Load partitioning executes portions of mobile's software on more than one device depending on energy and performance needs. Application-specific power optimizations have been implemented primarily for databases and multimedia. Most proxy adaptations to date have been relatively simple, such as dropping video content and delivering only audio in adverse conditions.

System level adaptation has been increasing in importance due to a large number of heterogeneous wireless devices present today. The mobiles themselves support multiple wireless interfaces, such as WLAN and GPRS. Mobility between the interfaces should happen seamlessly while still saving energy and meeting quality of service (QoS) needs. The next section outlines one such approach illustrated by a Hotspot scenario.

## 2. System-level power optimization example

Most of today's WLAN traffic is infrastructure oriented, and is routed through standard access points or more sophisticated Hotspots. When a new client enters the Hotspot environment it registers via an application level proxy before accessing data in the infrastructure. The data transmission then starts with no further power optimization on the part of the proxy.

This work extends the proxies with a resource manager. The resource manager's goal is to schedule data transmission times with clients in order to meet QoS requirements while minimizing the power consumption. The scheduler works similarly to the centralized WLAN low-power MAC schedulers or high data rate CDMA base station control algorithms, but in contrast occurs at a much higher level of abstraction and with the larger data burst sizes.

Larger data burst sizes mean that clients can have longer periods of sleep time, thus saving more energy. The quality of server's policies increases since it knows more about the clients in its network, such as their QoS needs, battery levels, current conditions in the channel etc. Client's resource manager can do a better job of determining application's future needs and implementing more sophisticated power management policies as it has access to information at all abstraction layers. On the other hand this approach performs best when resource manager middleware is present in all clients and the Hotspot server.

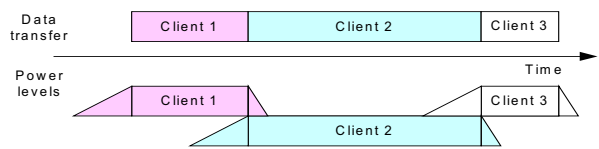

Figure 1. Sample schedule

Resource manager on the server dynamically selects the appropriate wireless network interface on each client (e.g. Bluetooth, WLAN), schedules data transfer in the large bursts of TCP or UDP packets and allocates appropriate bandwidth for communication. The client's resource manager implements the scheduling decisions by enabling data transfer and transitioning the wireless network interfaces (WNICs), between power states. It also aggregates information, such as its WLAN power state characteristics and QoS needs of the applications. Figure 1 shows a sample schedule. The top of the figure shows when data transfer occurs for each client. Power levels of clients are shown beneath. Since scheduling is centralized, each client knows exactly when it needs to wake up its WNIC and when it can enter a low power state.

A number of scheduling algorithms have been implemented in the Hotspot's [[3]] resource manager, ranging from standard real-time schedulers such as earliest deadline first, to well known packet level schedulers such as weighted fair queuing [[4]]. Figure 2 summarizes the average power measured for three concurrent IPAQ 3970 clients receiving high-quality MP3 audio first through standard WLAN and Bluetooth interfaces with no additional scheduling, and then with Hotspot scheduling. The scheduler initially has only Bluetooth enabled and as conditions in the link change, it seamlessly switches communication over to WLAN. In between delivery of larger bursts of data (10s of Kbytes at a time) the client's wireless devices enter low power modes: park for Bluetooth and off for WLAN (see Figure 1). QoS is maintained while saving 97% in WNIC power consumption.

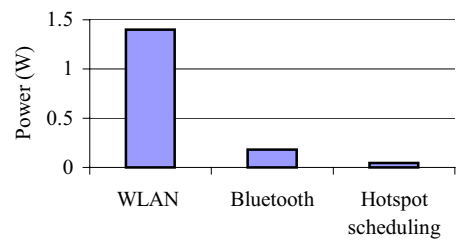

Figure 2. Average IPAQ power consumption

## 3. Conclusion

The need for energy savings in wireless communication will only increase as the demands for higher bandwidth access grow and battery lifetime continues to be a limitation. A number of power optimization techniques have been introduced in the past at all levels of WLAN design. System-level optimizations show a potential for even larger power savings obtained by utilizing methods available in all WLAN abstraction layers.

## 4. References

[1] C. E. Jones, K. M. Sivalingam, P. Agrawal, and J. C. Chen. "A survey of energy efficient network protocols for wireless networks", Wireless Networks, 7(4):343–358, July 2001.
[2] K. Holger, "An Overview of Energy-Efficiency Techniques for Mobile Communication Systems," TKN Technical Report, 2003.
[3] G. Manjunath, V. Krishnan, T. Simunic, J. Tourrilhes, A. McReynolds, D. Das, V. Srinivasmurthy, A. Srinivasan: "Smart Edge Server – going beyond access point," WMASH'04
[4] T. Simunic, W. Quadeer, G. De Micheli: "Managing heterogeneous wireless environments via Hotspot servers, " MMCN'05